\documentstyle[aps,preprint,psfig]{revtex}

\begin{document}
\title{A Stopped $\Delta$-Matter Source in Heavy Ion Collisions \\ at 10
GeV/$n$?}
\author{M.~Hofmann, R.~Mattiello, H.~Sorge,
H.~St\"ocker and
W.~Greiner}
\address{Institut f\"ur Theoretische Physik,
Universit\"at Frankfurt, 60054 Frankfurt, FRG}
\date{November 14th, 1994}
\maketitle
\begin{abstract}
We predict the formation of highly dense baryon-rich resonance matter in Au+Au
collisions at AGS energies. The final pion yields show observable signs for
resonance matter.
The $\Delta_{1232}$ resonance
is predicted to be the dominant source for pions of small transverse momenta.
Rescattering effects -- consecutive excitation and deexcitation of $\Delta$'s
-- lead to a long
apparent lifetime ($> 10\,{\rm fm}/c$) and rather large volumina
(several $100\,{\rm fm}^3$) of the $\Delta$-matter state.
Heavier baryon resonances prove to be crucial for reaction dynamics and
particle production at
AGS.
\end{abstract}


Many open questions in current nuclear and particle physics require high
energy and particle densities of nuclear matter to be resolved. The only
way to obtain these extreme conditions in the laboratory is the study
of relativistic heavy ion collisions \cite{STOECKER86}. In particular, it has
been shown that
the production of very high densities requires massive systems and high energy
as provided by the gold beam at AGS in Brookhaven
\cite{GONIN92}.
Several
in-medium effects are eventually present at high baryon densities, such as
resonance
matter \cite{BOGUTA81,CHAPLINE73}, mean fields due to changing quark and
gluon condensates \cite{COHEN92},
the decrease of baryon masses
due to chiral symmetry restoration \cite{ELLIS91} and the Quark-Gluon Plasma.
In particular, the formation of resonance matter contributes essentially to
enhanced strangeness \cite{MATT89} and subthreshold anti-proton production
\cite{SPIELES93}, baryon stopping and hadronic flow effects.

Microscopic and thermal calculations have shown that pion production
proceeds practically exclusively via the excitation of $\Delta$ resonances
in $AA$ reactions around 1 GeV/$n$ \cite{MISHUS,BASS94a}. However, since the
pion yield is much smaller than the number of participating nucleons, the
whole dynamics is dominated by nucleon interactions.

First efforts to investigate the $\Delta_{1232}$ production in
Si+Pb collisions the ultrarelativistic regime
directly
via $\pi N$ correlations have recently
been performed by the E814 Collaboration \cite{HEMMICK93,STACHEL93}.
In general, pions might be useful to learn about the
$\Delta$ resonance
\cite{SCHMIDT92,CASADO92,GYU79,TRZASKA91}
due to the almost sole decay mode
$\Delta\to N+\pi$ and the particular phase space distribution of $\pi$ from
$\Delta$-decays.

At higher energies ($E_{\rm lab}\approx 10$GeV),
a two-component structure in the transverse momentum spectra of pions
produced in $AA$ collisions has been observed \cite{AHMAD92,HEMMICK93}
which is not seen in elementary pp-reactions.

The low momentum component (about 50\%) is dominated by $\Delta$ decay as
suggested in \cite{BROWN91,SORGE91}. In Ref. \cite{SORGE91} the term
''$\Delta$ matter'' was coined to describe a system in which the $\Delta$
degree of freedom is as much excited as nucleons.
The second component in 10-15 GeV/$n$ collisions which is much stiffer than
the soft one has to be produced by another mechanism, possibly excitation of
higher mass baryonic resonances as will be elucidated in this paper.

Calculations with various transport models (QGSM, RQMD, hydrodynamics)
show a rich event shape differing markedly from the pure fireball
scenario. Collective flow seems to be strongly reflected in the baryonic
distributions (protons,
deuterons).
In order to take into account resonances as well as
flow effects and the complex event geometry, we use in this paper the RQMD
approach \cite{SORGE89a} to
predict the formation of dense resonance matter in violent heavy ion
reactions.
The ingredients of the model allow to extract explicitely the role
of resonance decays for the shape of the final $\pi$-spectra.

The RQMD model is a microscopic phasespace approach
in which the basic excitation mechanism is two-body particle scattering.
Both collision partners may get excited or they may annihilate into a single
$s$-channel state (e.~g. in meson-baroyn interactions). Several medium effects
have been explored in the RQMD approach, e.~g. mean fields \cite{MATTIELL94}
and string fusion
into color ropes \cite{SORGE92b}.
The $\Delta$-resonance is of special importance for the topics discussed in
this paper.
It should be noted that $\Delta$ production (e.~g. in $NN\to N\Delta$),
absorption
(e.~g. $\Delta N\to NN$ and $\pi\Delta\to B^*$), formation ($\pi N\to\Delta$)
and
the decay ($\Delta\to\pi N$) are treated dynamically. We use non-fixed
$\Delta$-masses, improved detailed balance relations for time-reversed cross
sections and
the short lived nature of the $\Delta$-resonance.
Low mass excitations (e.~g. nonstrange baryons with mass $<2 {\rm GeV}/c^2$)
are projected on discrete resonance states, higher mass excitations decay
stringlike \cite{ANDER82}.
The model describes well nuclear stopping and particle production from 1 to
200 GeV/$n$
\cite{HOFMANN93,SORGE91}.
Furthermore, the source size extracted from RQMD events with the HBT
technique agrees well
with recent measurements \cite{NUXU93}.


Fig.\ref{fig2} shows the transverse mass spectrum of freeze-out
$\pi^-$ in Au(10.7 GeV/$n$)Au with central impact parameter ($b\le3$
fm) and potential interaction included. The solid line indicates the total
yield whereas the other
curves represent the contribution from various final resonance decays.
Other pion sources (elastic MM or MB scattering) are neglected.

The integrated spectrum shows a bending
that is suggestive of two different contributions. The calculations give that
at
low transverse momenta most of the pions stem
from $\Delta$ decays.
Pions
from heavy resonances contribute essentially at high $p_t$ and
dominate the spectrum for $m_t-m_0 > 600$ MeV. This behaviour is quite easy
to understand because the decay of highly excited resonances into pions will
produce decay products with more kinetic
energy than the decay of a $\Delta$.
$\rho$ mesons
contribute almost 10\% of the total yield, higher resonances only to
5\%.
The decays of $\eta$, $\eta'$, $\omega$ or $K^*$ into pions and the pion
elastic scattering add up to another 15\% of the total yield that will be
neglected in our consideration.

The upper part of Fig. \ref{siginv} shows the $\Delta$-source producing most
of
the final pions revealing a strong peak at midrapidity.
Experimentally the dominance of $\Delta$-resonance decays for low-$p_t$ pions
may be verified
from the single spectra using a technique
proposed by one of us \cite{SORGE93}.
If we turn to rapidities far from midrapidity the slopes of the distribution
will change
drastically with $y$. If the transverse momentum of a pion produced
by $\Delta$-decay vanishes ($p_t\approx 0$), we will obtain a
longitudinal momentum $p_z = 227 \mbox{MeV}/c$ in the $\Delta$ rest frame,
(assuming a fixed $\Delta$ mass $m_{\Delta} = 1232 {\rm MeV}/c^2$). This
corresponds
to a rapidity of 1.27. Thus the pion is fed into a completely different
rapidity region.
At rapidities with highest densities
of $\Delta$ resonances we therefore expect a suppression of ''ultrasoft''
($p_t\to 0$) pions.
The effect just described becomes visible by plotting the ratio of the
production
cross section for ''ultrasoft'' pions with $p_t\approx 0$ and ''soft'' pions
with $p_t\approx 150 {\rm MeV}$.
In the lower part of Fig. \ref{siginv} the ratio of invariant production
cross sections for $\pi^+$
\begin{equation}
C_y = \frac{\sigma_{\rm inv}(p_t = 30 {\rm MeV})}{\sigma_{\rm inv}(p_t = 150
{\rm MeV})}
\end{equation}
is shown versus rapidity for central ($b<3{\rm fm}$) Au+Au at 10.7 GeV/$n$.
The distribution shows a minimum at $y=0$: the same rapidity where
the stopped $\Delta$ source is located.
Within a surprisingly small relative momentum bin ($-1\le y\le +1$)
we find 189 $\Delta$'s (48\% of all baryons) so that it is reasonable to
speak about
a $\Delta$-matter state at $y=0$.

Since we obtain that huge amount of resonances at 10 GeV/$n$ contributing to
the final spectra, we are now interested in the dynamics of the resonances
during
the reaction.
Fig. \ref{Fig1} shows the time evolution of the number of resonances in
central
Au+Au reactions ($b<3$ fm) at 1 GeV/$n$ (left) and 10.7 GeV/$n$ (right).
Time is always taken in the center-of-momentum frame of the whole reaction.
The thick solid line represents the sum of all particles in the system except
nucleons. The solid line shows the nucleons while the dashed and dotted lines
show the $\Delta_{1232}$ and all baryon resonances, respectively.


At AGS, between 4 and 7 fm/$c$
the number of baryon resonances in the system exceeds the number of nucleons.
In the early hot and dense stages of the reaction
the main part of total particle production is
performed (thick solid line). With increasing time the system becomes more
and more equilibrated and the excited resonances in the system consecutively
evaporate
mesons.
In this picture the
lightest baryon resonance state, the $\Delta_{1232}$ (dashed line), proves to
be a very
important and interesting state. At the point of maximal
excitation the $\Delta$ contributes nearly 50\% to all resonances.
With increasing time, only the $\Delta$ survives:
its apparent lifetime ($\tau=15{\rm fm}/c$) is much longer than that of the
other
excited states due to consecutive $\Delta\to\pi N,\pi N\to\Delta$ processes
that
are the dominant reactions in the final stage \cite{HOFMANN94b}. This
becomes visible by switching
off the $\Delta$ formation channel $\pi N\to\Delta$ which yields a decrease
of the
$\Delta$-lifetime to $\approx 5{\rm fm}/c$.


In order to predict a resonance {\em matter} state we have to check
whether the density of resonances and $\Delta$'s is comparable to nuclear
ground state
density.
Furthermore, we
demand the resonances to reach these densities in a rather large volume of
the same
order of magnitude as the reaction zone. For this, in Fig. \ref{dichte}
the time evolution of densities and occupied volume of baryons (solid line)
and $\Delta$'s (dashed) is plotted. The reaction zone is divided into small
boxes of $\Delta x\Delta
y\Delta z = 1 \mbox{fm}\times 1\mbox{fm}\times 2 \mbox{fm}$.
The density is calculated by averaging over the 10\% boxes with the highest
density. To
obtain the volume all boxes with densities larger than 
$0.1\,\rho_0$ are added up.
The so defined mean baryon denstity reaches 7 times nuclear matter density.
The density of $\Delta$ resonances also exceeds nuclear density and reaches
2.5$\rho_0$.
At the time of maximum $\Delta$ density the $\Delta$ volume becomes
$V_{\Delta}=500 {\rm fm}^3\approx\frac{1}{3}\,V_B$
being in the same order of magnitude as the reaction volume.
The average baryon and $\Delta$-densities in this volume are 3$\rho_0$ and
$\rho_0$, respectively.
Up to 10 fm/$c$ the $\Delta$ density is still $\ge\rho_0$ while the volume
becomes
$V_{\Delta}\approx 300{\rm fm}^3$.



In order to extrapolate our results to other energy regions we plot in Fig.
\ref{Fig3} the resonance excitation
function for $^{83}$Kr+$^{83}$Kr without potentials
for $\Delta_{1232}$ (circle), $N^*_{1520}$ (cross), $\Delta_{1600}$ (dagger)
and $\Delta_{1700}$ (asterisk). The integrated number of all resonances is
given by the dotted curve (triangle).
The solid line (square) shows the number of nucleons. The thick solid
line stands for the maximum number of all particles except nucleons.
The values are taken at the time
of maximum number of baryon resonances in the system.
One observes that the number of excited resonances increases with beam energy.
At energies $\ge$10 GeV/$n$ there
are more resonances than nucleons in the system even in
a light system as Kr+Kr. Higher energies cause further but much slower
increase
of the resonance number, while lower energies reveal a fast fall off.

Recent claims of resonance matter formation at GSI-SIS energies
\cite{METAG93} are scrutinized
in Fig. \ref{Fig1} (left).
Only 10--15\% of all baryons are excited to resonances,
most of them
$\Delta_{1232}$, so that a description without higher resonances might be
possible.
But due to the smaller number
of resonances nucleonic degrees of freedom are most prominent. It would be
an euphemism to call such a state resonance matter \cite{BASS94a}.
The higher the energy the more
heavy resonances as $N^*$ or $\Delta^*$ are produced,
but even at CERN energies (200 GeV/$n$) half of all resonances are
$\Delta_{1232}$.
Thus the $\Delta$ remains the most important resonance state over a wide
energy range, but for energies higher than 2 GeV it is necessary to
take the dynamics of higher mass resonances into account.


\section*{Conclusion}

Within the RQMD approach we predict at AGS energies
the existence of a dense resonance matter
state in a large volume during the early part of the reaction. Particularly,
the $\Delta$-matter state reaches densities of $2.5 \rho_0$.
Therefore, the investigation of
the $\Delta$ propagation in a dense medium \cite{TERHAAR87}
is of most importance for an understanding of the $AA$ reactions at
1 to 20 GeV$/n$.

Higher mass baryon resonances contribute to the final pion spectra
preferentially at high $p_t$,
where they clearly exceed the $\Delta$'s contribution. Even in absolute
numbers
at early stages of the reaction, neglecting higher resonances leads to an
overpopulation of the $\Delta_{1232}$ compared to our results. The $\Delta$
proves to be the dominant state and main source for pion production
at low transverse momenta $p_t<300$MeV. This promises a
way to get experimental information on $\Delta$-matter by analyzing
especially the low-$p_t$ pions in order to localize
$\Delta$-matter in momentum space. Experimental research in near future
should be able to confirm this prediction.
However, even at energies of 1 GeV/$n$ it has been shown
\cite{MATT89,SPIELES93}
that higher mass resonances
are crucial to understand $K$, $\bar p$ and other heavy particle yields via
multi-step processes.

\begin{figure}
\caption{Transverse mass spectra of $\pi^-$ in Au+Au at 10.7 GeV/$n$, $b\le
3$ fm. Total yield (solid line), pions produced by decay of
$\Delta_{1232}$ (dashed), $\rho$ (dashed-dotted) and other baryons and
strings (dotted). At low transverse momenta the $\Delta$ decay
dominates all the other pion producing channels. For higher $p_t$ the
other baryon resonances' contribution grows and finally dominates the
spectrum. The bending of the total distribution is caused by a superposition
of the different shapes of the $\Delta$
constribution at low and the baryon distribution at high $p_t$.
Thus, low $p_t$ pions might provide a useful tool to extract information
about the properties of $\Delta$ matter.}
\label{fig2}
\end{figure}

\begin{figure}
\caption{Upper part: Rapidity distribution of $\Delta$'s in Au+Au
at 10.7 GeV/$n$. Only $\Delta$'s whose decay leads to a freeze-out pion are
considered.
There are 189 $\Delta$'s within the rapidity range $-1\le y\le +1$ so that we
find
$\Delta$-matter at midrapidity. Lower part: Ratio of invariant cross section
of $\pi^+$ vs. rapidity
for $p_t=30\pm 20$ MeV and
$p_t=150\pm 20$ MeV. A strong minimum at midrapidity and therefore the
existence of a
$\Delta$-matter state is revealed.}
\label{siginv}
\end{figure}

\begin{figure}
\caption{Time evolution of particles in central Au+Au collisions at 1 and
10.7 GeV/$n$,
representing SIS and AGS energy ranges.
Plotted are nucleons (solid line), baryonic resonances (dotted),
$\Delta_{1232}$
(dashed) and all particles in the system that are no nucleons (thick solid).
Left part: At SIS, only 10\% of the nucleons are excited into resonances.
Most of the resonances
are $\Delta_{1232}$. Thus, there is no
hint for a resonance matter state at SIS energies.
Right part: At AGS there are more excited baryons than nucleons in the system.
The $\Delta$ resonance only contributes
about 50 percent to the resonance matter. With increasing time the $\Delta$
becomes more and more important until most of the baryonic resonances are
$\Delta$'s. Thus, the $\Delta$ resonance seems to decay much slower than other
resonances. This behaviour is due to the large $\pi N\rightarrow \Delta$ cross
section that causes
a consecutive excitation and deexcitation of $\Delta$'s.}
\label{Fig1}
\end{figure}

\begin{figure}
\caption{
Upper part: Time evolution of baryon and $\Delta$ densities. The figure shows
the averaged
densities over the 10\% densest coordinate space boxes. The boxsize is chosen
to be $\Delta x \Delta y \Delta z = 1$fm$\times 1$fm$\times 2$fm.
The density of all baryons reaches seven times nuclear density while
the density of $\Delta$ resonances is 2.5$\rho_0$. One observes
high numbers of resonances at densities higher than nuclear matter density.
\quad Lower part: Volumina of baryon and $\Delta$ resonances. Only boxes with
densities larger
than $0.1\,\rho_0$ are considered.
}
\label{dichte}
\end{figure}

\begin{figure}
\caption{Maximum number of nucleon resonances divided by total baryon number
in relation to bombarding energy in central $^{83}$Kr+$^{83}$Kr collisions
($b<2$)
calculated with RQMD in cascade mode. Calculations were done
at 1, 1.5, 2, 5, 10 and 200 GeV.
One easily sees the domination of the $\Delta_{1232}$ resonance in
all energy ranges, although the relative contribution of the higher resonances
increases with bombarding energy. Beam energies above 10 GeV show
more baryonic resonances than nucleons in the system.}
\label{Fig3}
\end{figure}
\pagebreak
\pagestyle{empty}

{\Huge Figure 1:}\\[3cm]
\centerline{\psfig{figure=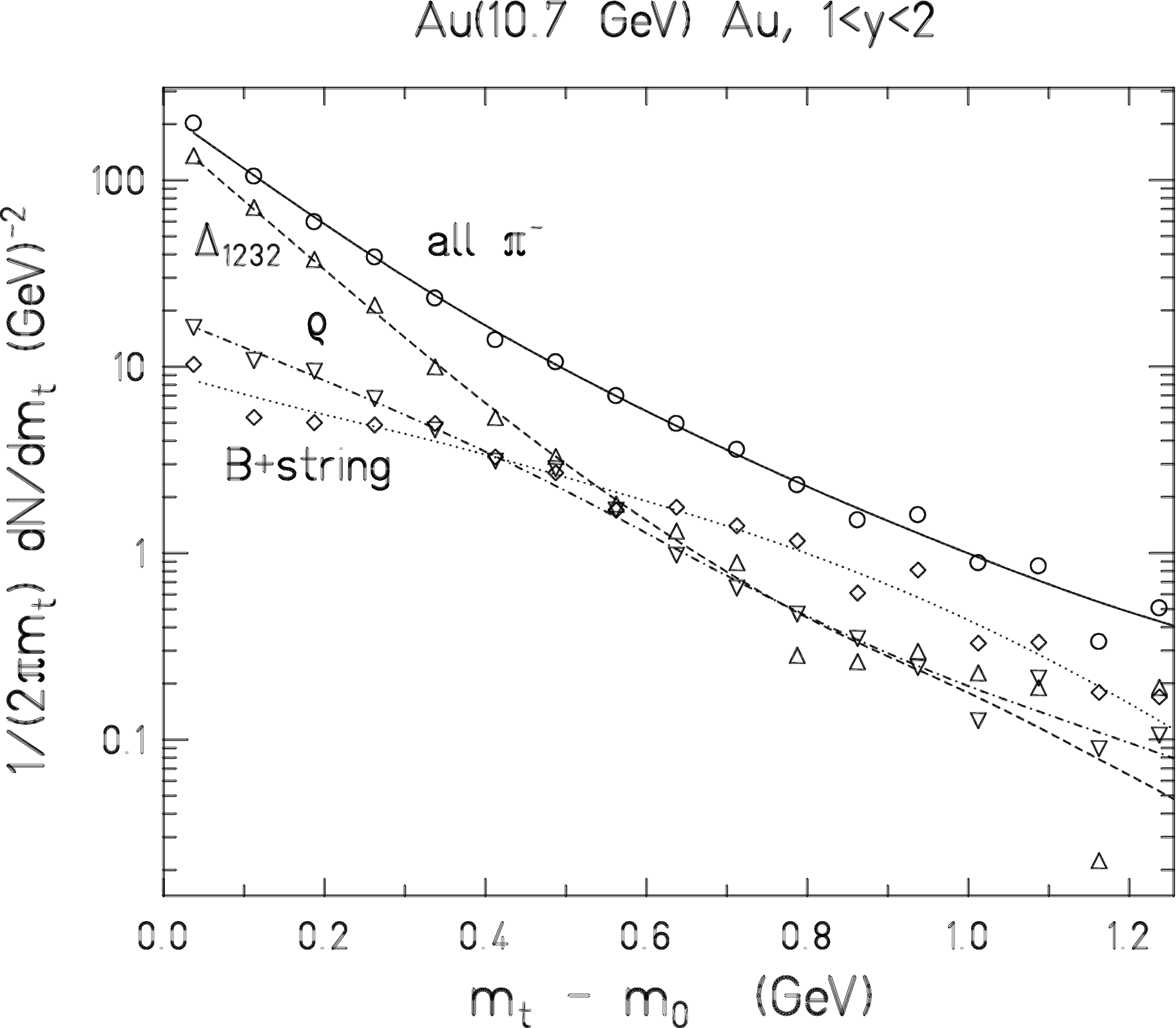,width=16cm}}
\pagebreak

{\Huge Figure 2:}\\[2cm]
\centerline{\psfig{figure=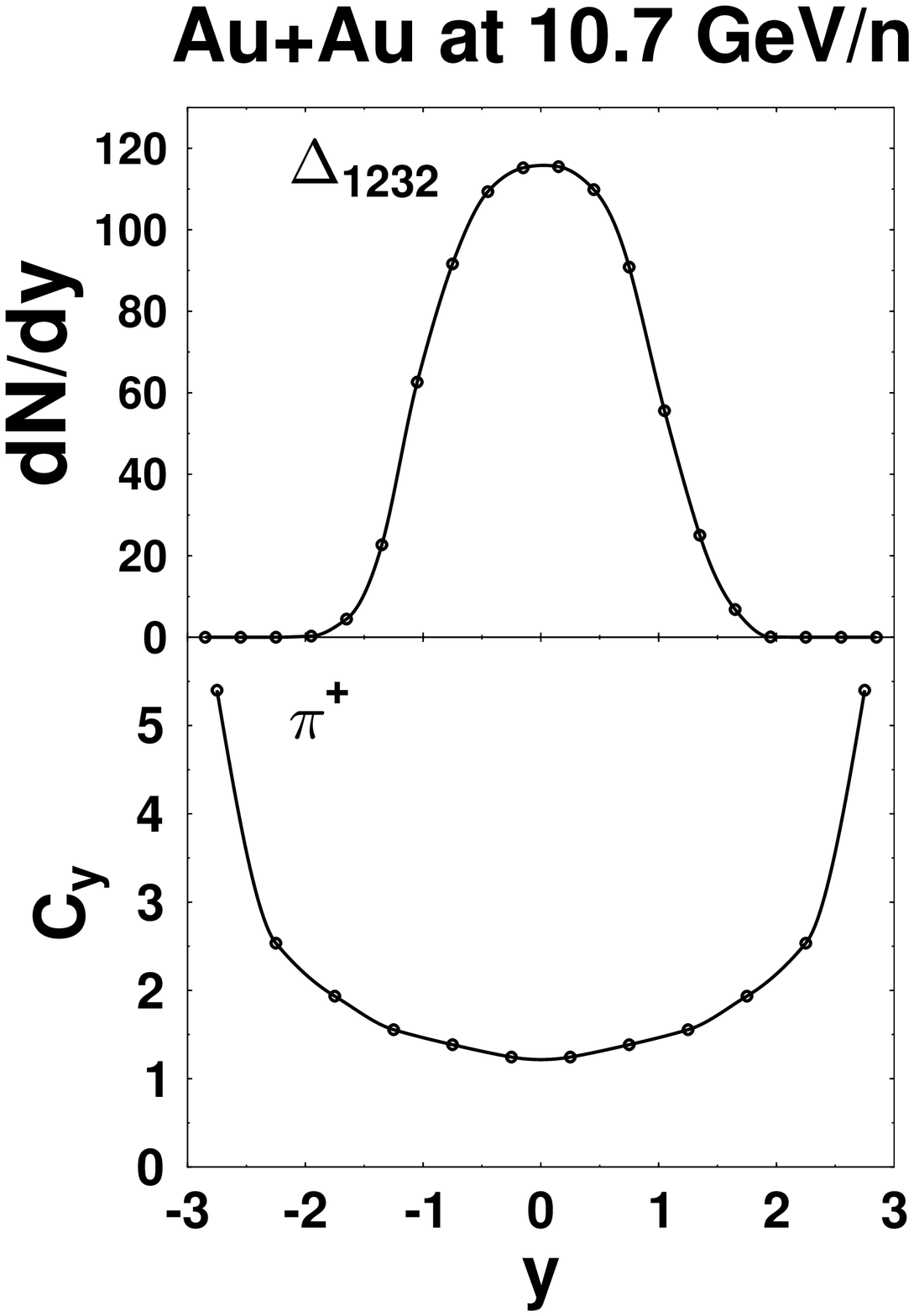,width=14cm}}
\pagebreak

{\Huge Figure 3:}\\[3cm]
\centerline{\psfig{figure=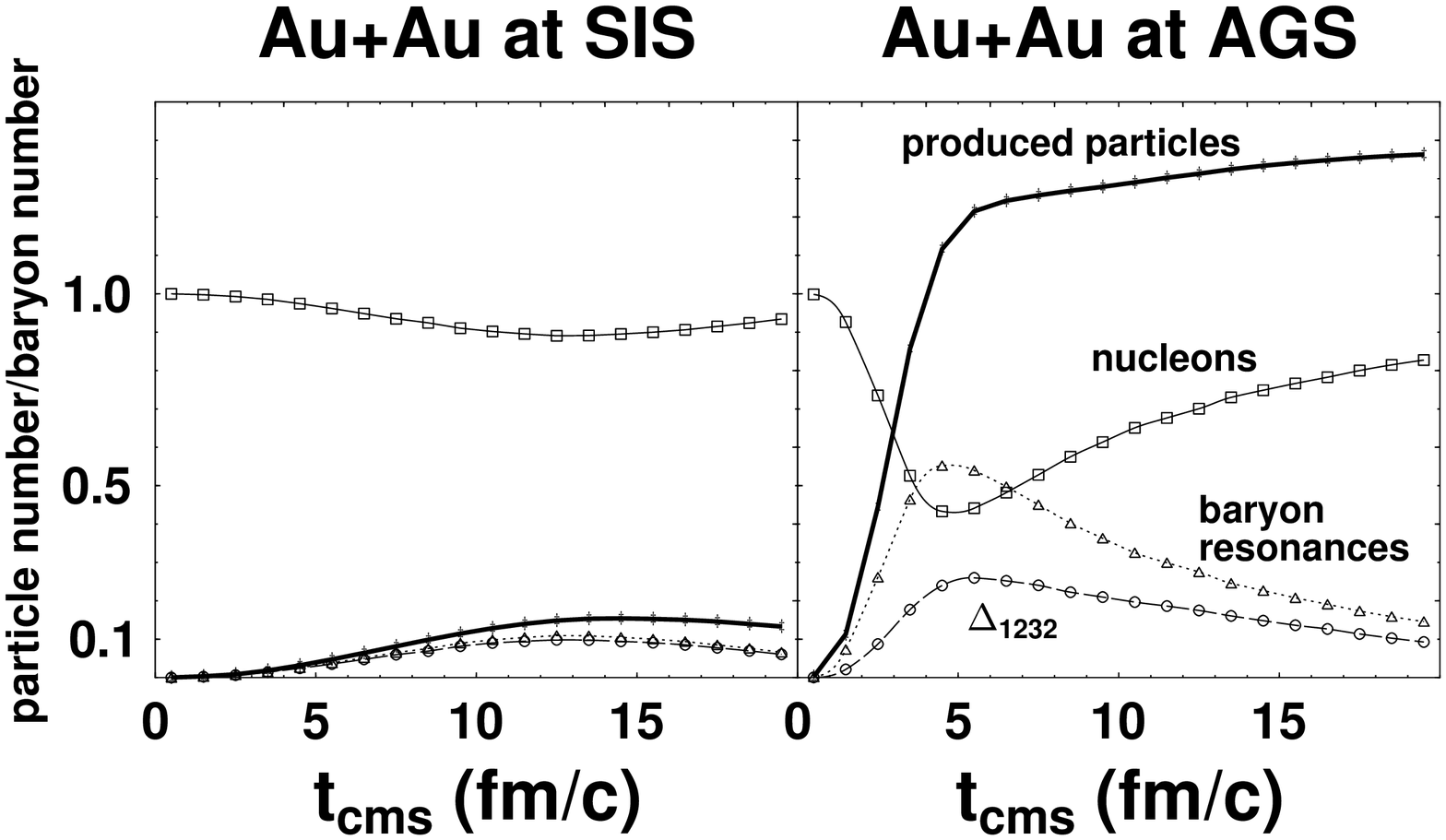,width=16cm}}
\pagebreak

{\Huge Figure 4:}\\[3cm]
\centerline{\hspace{3cm}\psfig{figure=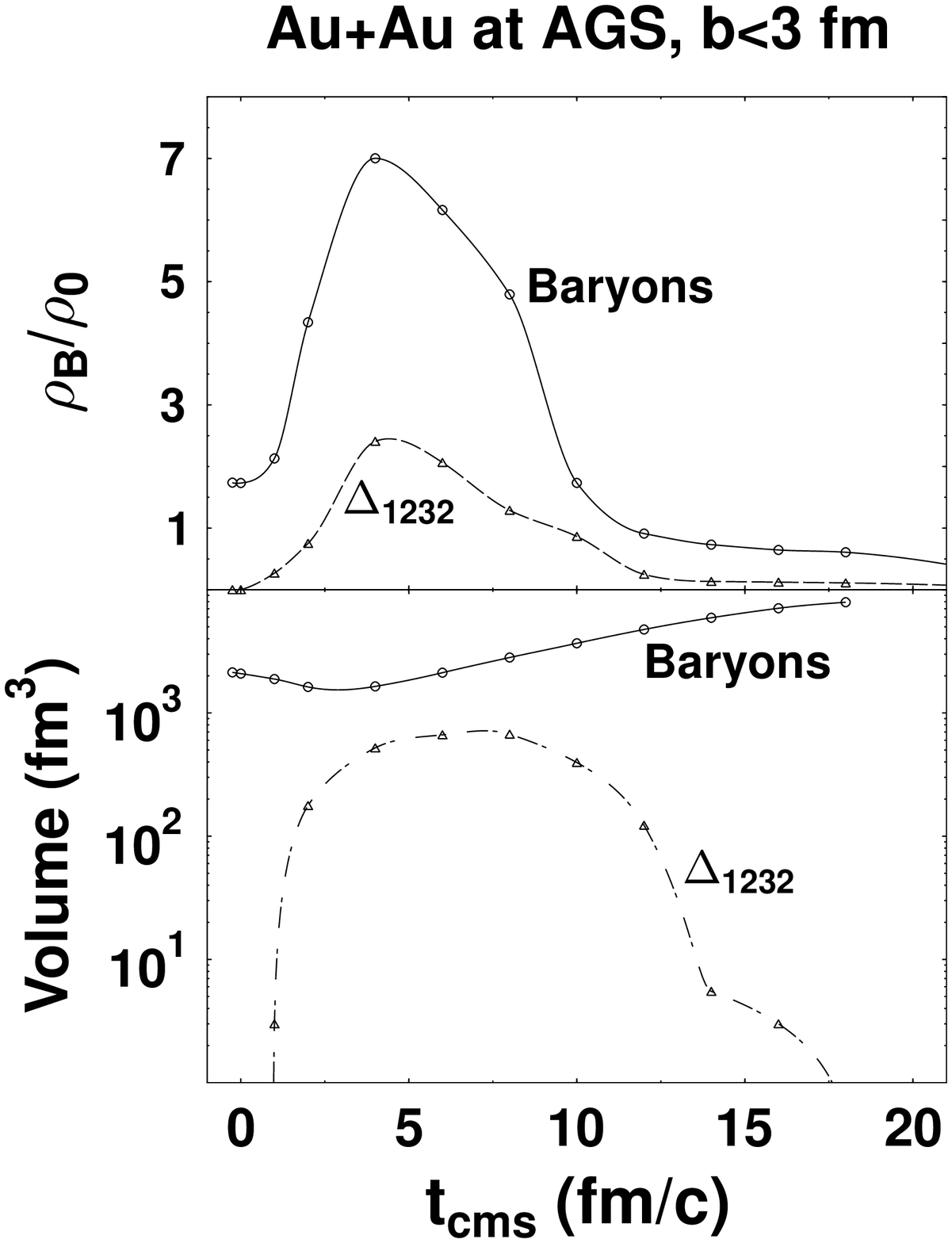,height=18cm}}
\pagebreak

{\Huge Figure 5:}\\[3cm]
\centerline{\psfig{figure=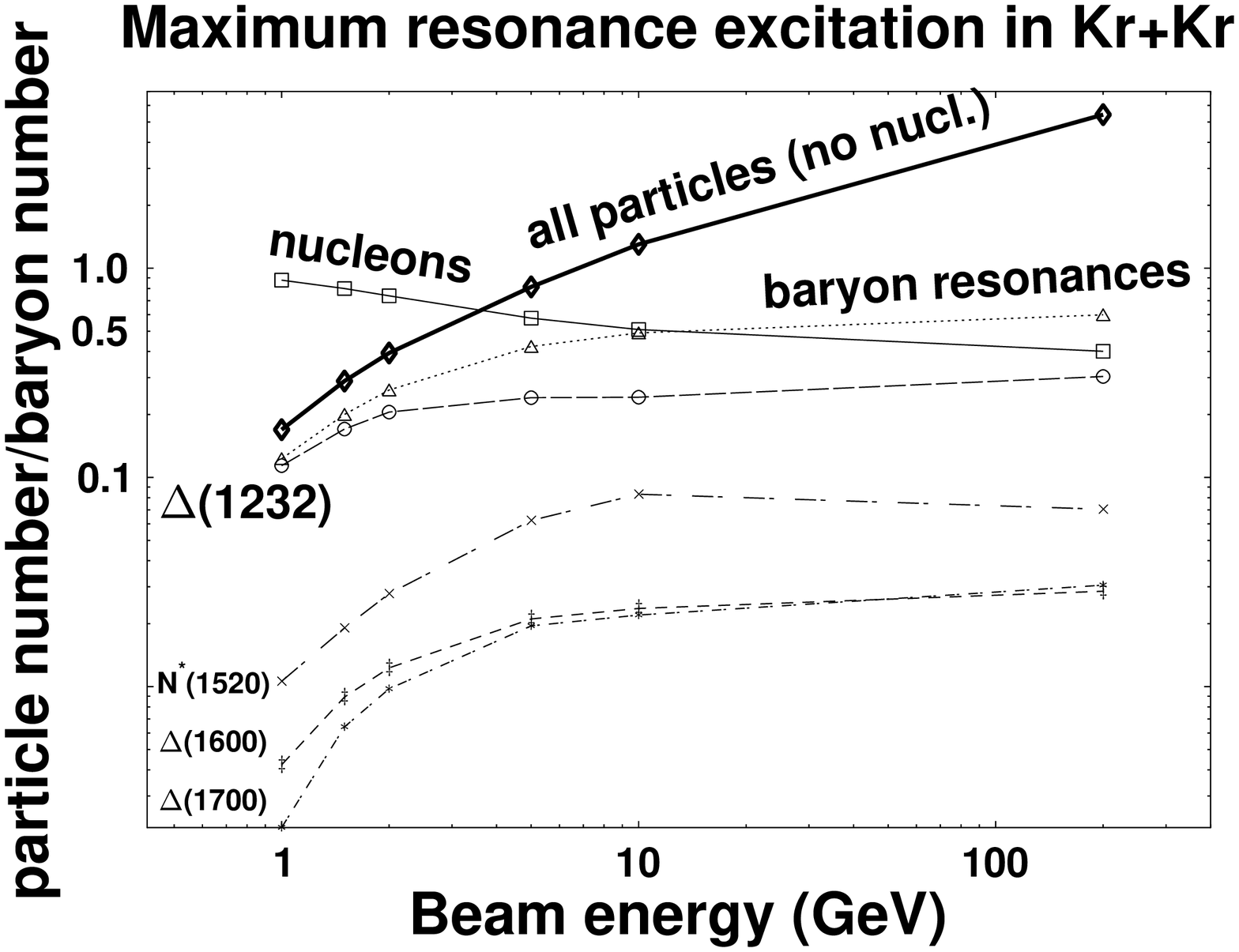,width=16cm}}

\end{document}